\documentclass[manuscript,screen,nonacm]{acmart}

\AtBeginDocument{%
  \providecommand\BibTeX{{%
    \normalfont B\kern-0.5em{\scshape i\kern-0.25em b}\kern-0.8em\TeX}}}

\setcopyright{acmcopyright}
\copyrightyear{2018}
\acmYear{2018}
\acmDOI{XXXXXXX.XXXXXXX}

\acmConference[Conference acronym 'XX]{Make sure to enter the correct
  conference title from your rights confirmation emai}{June 03--05,
  2018}{Woodstock, NY}
\acmPrice{15.00}
\acmISBN{978-1-4503-XXXX-X/18/06}

\usepackage{array}
\newcolumntype{L}[1]{>{\raggedright\let\newline\\\arraybackslash\hspace{2pt}}m{#1}}
\newcolumntype{C}[1]{>{\centering\let\newline\\\arraybackslash\hspace{0pt}}m{#1}}
\newcolumntype{R}[1]{>{\raggedleft\let\newline\\\arraybackslash\hspace{0pt}}m{#1}}
\usepackage{float}
\usepackage{multirow}
\usepackage{makecell}
\usepackage{booktabs}

\usepackage{graphicx}
\usepackage{subfig}

\DeclareMathOperator{\softmax}{Softmax}
\DeclareMathOperator{\relu}{ReLU}

\usepackage{bm}  

\usepackage{enumitem}
\newlist{research_questions}{enumerate}{1}
\setlist[research_questions]{label*=\textbf{RQ\arabic*}}

\usepackage{xcolor}
\usepackage[draft,inline,nomargin,index]{fixme}

\definecolor{darkgreen}{RGB}{0,100,0}
\fxsetup{theme=color,mode=multiuser}
\FXRegisterAuthor{pranav}{pranavK}{\colorbox{darkgreen}{\color{white}Pranav}}
\FXRegisterAuthor{elias}{Elias}{\colorbox{blue}{\color{white}Elias}}
\FXRegisterAuthor{prof}{Prof}{\colorbox{red}{\color{white}prof. Pasi}}
\FXRegisterAuthor{rev}{reviewer}{\colorbox{orange}{\color{white}Reviewer}}

\usepackage{soul}

\begin{document}

\title{Denoising Attention for Query-aware User Modeling in Personalized Search}


\author{Elias Bassani}
\email{e.bassani3@campus.unimib.it}
\orcid{0000-0001-7922-2578}
\affiliation{%
  \institution{Consorzio per il Trasferimento Tecnologico - C2T}
  \city{Milan}
  \country{Italy}
}
\affiliation{%
  \institution{University of Milano-Bicocca}
  \city{Milan}
  \country{Italy}
}

\author{Pranav Kasela}
\email{p.kasela@campus.unimib.it}
\orcid{0000-0003-0972-2424}
\affiliation{%
  \institution{University of Milano-Bicocca}
  \city{Milan}
  \country{Italy}
}

\author{Gabriella Pasi}
\email{gabriella.pasi@unimib.it}
\orcid{0000-0002-6080-8170}
\affiliation{%
  \institution{University of Milano-Bicocca}
  \city{Milan}
  \country{Italy}
}
\renewcommand{\shortauthors}{Bassani, et al.}

\begin{abstract}
The personalization of search results has gained increasing attention in the past few years, thanks to the development of Neural Networks-based approaches for Information Retrieval and the importance of personalization in many search scenarios.
Recent works have proposed to build user models at query time by leveraging the \textit{Attention} mechanism, which allows weighing the contribution of the user-related information w.r.t. the current query.
This approach allows taking into account the diversity of the user's interests by giving more importance to those related to the current search performed by the user. 

In this paper, we first discuss some shortcomings of the standard \textit{Attention} formulation when employed for personalization.
In particular, we focus on issues related to its normalization mechanism and its inability to \textit{entirely} filter out noisy user-related information.
Then, we introduce the \textit{Denoising Attention} mechanism: an \textit{Attention} variant that directly tackles the above shortcomings by adopting a robust normalization scheme and introducing a filtering mechanism.
The reported experimental evaluation shows the benefits of the proposed approach over other \textit{Attention}-based variants.
\end{abstract}





\maketitle


\section{Introduction}\label{intro}

The past few years have witnessed an increasing interest in the application of Deep Learning techniques for tackling various tasks of Information Retrieval \cite{DBLP:journals/ipm/GuoFPYAZWCC20}, such as Personalized Search.

Two of the main challenges of \textit{Personalized Search} are \textit{how} and \textit{when} personalization should take place.
First, not all the data gathered to represent specific users' preferences in user models are equally related to each of the user's searches, as users usually have multiple and diverse interests.
Second, personalization is not always beneficial to the retrieval process \cite{DBLP:conf/sigir/TeevanDL08} as it could cause the information need expressed by the user to be misinterpreted by the system.
For example, this might happen when the user's interests in a specific domain are \textit{unknown}, but the system still applies personalization, which, in this case, leverages user-related information \textit{potentially unrelated} to the user's information need; this could ultimately decrease the system's effectiveness.

A recent trend in \textit{Personalized Search} \cite{DBLP:conf/cikm/GeDJNW18,DBLP:conf/sigir/LuDXWW20,DBLP:conf/wsdm/ZhouDW20,DBLP:conf/cikm/AiHVC19,DBLP:conf/sigir/ZhongGGL20,DBLP:conf/www/YaoDXW20,DBLP:conf/www/JiangWRCYCGJC020,DBLP:conf/sigir/ZhouDW20,DBLP:conf/sigir/BiAC20,DBLP:conf/sigir/BiAC21,DBLP:conf/sigir/LuDJNW19} is \textit{query-aware user modeling}, which consists in building a representation of the user preferences, \textit{i.e.}, the user model, at query time, based on various sources of user interests and by giving more importance to those related to the current search performed by the user.
Since a user is typically interested in different and even \textit{unrelated} topics, a desirable property for defining reliable personalization models is the ability to discern between beneficial and noisy user-related information on a query basis.
Previous works in this context that make use of neural models rely on the \textit{Attention} mechanism \cite{DBLP:journals/corr/BahdanauCB14},
to differently weigh the contribution of distinct sources of user-related information in building the user representation at query time.
Despite the increasing use of the \textit{Attention} mechanism in user modeling, there is still a lack of an in-depth analysis of its behavior and effects on personalization, as well as a systematic comparison with simpler operators in this context.

In this paper, we first describe and analyze the \textit{Attention} mechanism when used for query-aware user modeling, by highlighting some shortcomings of the standard \textit{Attention} formulation, related to its use of the \textit{Softmax} function (Section~\ref{preliminaries}).
Specifically, the \textit{exponential} function employed by the \textit{Softmax} can cause the user model to be excessively noisy or skewed towards a single piece of user-related information.
Moreover, as it will be extensively discussed in Section ~\ref{preliminaries}, due to the fact that the standard \textit{Attention} mechanism uses the \textit{Softmax}'s outputs to weigh the contribution of the sources of user information to build the user model at query time, personalization is performed even when those sources are not related to the current search conducted by the user.
To overcome these weaknesses, in Section ~\ref{denoising}, we propose the \textit{Denoising Attention} mechanism, an \textit{Attention} variant specifically designed to finely filter out noisy user-related information and to produce a balanced representation of the user interests w.r.t. the current search.
Firstly, we introduce a novel filtering mechanism based on the \textit{Rectifier Linear Unit} \cite{DBLP:conf/icml/NairH10} and a threshold value. Secondly, we depart from the \textit{Softmax} function and opt for a more straightforward and robust weighting scheme.
To evaluate our proposal, we tackle the task of \textit{Personalized Results Re-Ranking}; to make a comparative evaluation of the proposed user model with alternative user models at the state-of-the-art, we rely on a framework that allows us to switch the user representation model with ease. We introduce the considered task and the related framework in Section~\ref{re-ranking}. 
Then, we present the research questions we addressed and describe the experimental setup of our comparative evaluation (Section~\ref{experimental_setup}).
Finally, in Section~\ref{eval}, we present the comparison between the \textit{Denoising Attention}, the standard \textit{Attention}, the \textit{Zero Attention strategy} \cite{DBLP:conf/cikm/AiHVC19}, an \textit{Attention} variant previously proposed for user modeling, and the \textit{Multi-Head Attention} \cite{DBLP:conf/nips/VaswaniSPUJGKP17}, and we also ablate our proposed \textit{Attention} variant.
The results of our evaluation clearly show the advantages of \textit{Denoising Attention} and the importance of the filtering mechanism it implements.
We share all the code to reproduce the experimental evaluation, and we make available the implementation of \textit{Denoising Attention} for future works\footnote{We will add a link to the repository upon acceptance}.

\section{Related Work}\label{related_works}
Personalization of search results has received considerable attention from both academia \cite{DBLP:conf/ecir/VuNJSW17,DBLP:conf/cikm/GeDJNW18,DBLP:conf/sigir/LuDXWW20,DBLP:conf/www/YaoDXW20,DBLP:conf/sigir/ZhouDW20,DBLP:conf/sigir/BiAC20,DBLP:conf/sigir/BiAC21,DBLP:conf/sigir/LuDJNW19,DBLP:conf/cikm/CarmanCHB10,DBLP:conf/cikm/SiegMB07,DBLP:conf/kdd/TanSZ06,DBLP:conf/cikm/HarveyCC13,DBLP:conf/ictai/PretschnerG99,DBLP:conf/wsdm/ZhouDW20,DBLP:journals/sigir/KellyT03} and industry \cite{DBLP:conf/recsys/Grbovic17,DBLP:conf/kdd/GrbovicC18,DBLP:conf/wsdm/SongWH14,DBLP:conf/sigir/ZhangWZTJXYY20,li2014deep,DBLP:conf/sigir/BennettWCDBBC12,DBLP:conf/www/DouSW07,DBLP:conf/wsdm/SontagCBWDB12,DBLP:conf/wsdm/MatthijsR11,DBLP:conf/sigir/XuBFSY08,DBLP:conf/cikm/CarmelZGOHRUYC09,DBLP:conf/sigir/TeevanDL08,DBLP:conf/cikm/AiHVC19,DBLP:conf/sigir/ZhongGGL20,DBLP:conf/www/JiangWRCYCGJC020,DBLP:conf/sigir/RadlinskiD06,DBLP:conf/sigir/TeevanDH05}. 
Over time, many different approaches have been proposed to tackle this task.
The definition of user models is a core issue in the task personalized search. 
To the aim of defining reliable user models, early approaches
relied on
click-based features \cite{DBLP:conf/sigir/BennettWCDBBC12,DBLP:conf/www/DouSW07,DBLP:conf/sigir/TeevanDL08,DBLP:conf/wsdm/TeevanLG11},
content-based features \cite{DBLP:conf/wsdm/MatthijsR11,DBLP:conf/sigir/TeevanDL08},
social network analysis \cite{DBLP:conf/cikm/CarmelZGOHRUYC09},
language models \cite{DBLP:conf/kdd/TanSZ06,DBLP:conf/wsdm/SontagCBWDB12},
topic modeling \cite{DBLP:conf/cikm/HarveyCC13,DBLP:conf/cikm/CarmanCHB10,DBLP:conf/sigir/XuBFSY08},
ontologies \cite{DBLP:conf/cikm/SiegMB07,DBLP:conf/ictai/PretschnerG99},
and other sources of user-related information as well as other formal means to build user representations.
Researchers have focused on the application of Deep Learning and Word Embeddings for the personalization of search results \cite{li2014deep,DBLP:conf/wsdm/SongWH14,DBLP:conf/sigir/ZhangWZTJXYY20,DBLP:conf/ecir/VuNJSW17}.
These works take advantage of the opportunity given by Representation Learning \cite{DBLP:journals/pami/BengioCV13} to build latent semantic vector representations of queries, documents, and the gathered user-related information.
Recently, a new trend in modeling user interests has emerged, which aims at building a representation of the user preferences, i.e., the user model, at query time.
In particular, several works \cite{DBLP:conf/cikm/GeDJNW18,DBLP:conf/sigir/LuDXWW20,DBLP:conf/wsdm/ZhouDW20,DBLP:conf/cikm/AiHVC19,DBLP:conf/sigir/ZhongGGL20,DBLP:conf/www/YaoDXW20,DBLP:conf/sigir/ZhouDW20,DBLP:conf/www/JiangWRCYCGJC020,DBLP:conf/sigir/BiAC20,DBLP:conf/sigir/BiAC21} rely on the \textit{Attention} mechanism to weigh and aggregate the available user-related information on a query basis. 
With the aim of applying query-aware personalization, these models try to take advantage of the diverse interests that a user may have.
Many previous works \cite{DBLP:conf/cikm/GeDJNW18,DBLP:conf/sigir/LuDXWW20,DBLP:conf/www/YaoDXW20} rely on the \textit{Attention} mechanism to weigh, w.r.t. the current query, the contribution of a user's prior search sessions (represented as a combination of the representations of the previous queries and those of the documents accessed by the user after issuing them) for composing the user model employed for conducting session-based personalization.
Zhou et al.~\cite{DBLP:conf/wsdm/ZhouDW20} tackle the problem of user re-finding behavior and rely on \textit{Attention} to weigh, w.r.t. a user's current search, the previous queries she issued, and the documents she accessed in the past, which are then used as sources of the user interests for personalizing the current search results.
By leveraging the \textit{Attention} mechanism, Zhong et al.~\cite{DBLP:conf/sigir/ZhongGGL20} weigh user-related terms w.r.t. the query the user is typing, and use a weighted combination of those terms and the original query terms to generate personalized query suggestions.
Jiang et al.~\cite{DBLP:conf/www/JiangWRCYCGJC020} propose an \textit{attentive} Personalized Item Retrieval model leveraging the \textit{Attention} mechanism to estimate the importance of each item in the user history while conducting personalization.
Despite the increasing application of the \textit{Attention} mechanism for user modeling, the vast majority of previous works do not conduct an in-depth analysis of its behavior and effects on personalization.
The sole exception is represented by the \textit{Zero Attention Model} proposed by Ai et al.~\cite{DBLP:conf/cikm/AiHVC19}. The authors propose an \textit{Attention} variant defined to allow the retrieval model to avoid personalization when no source of user information is related to her current search, which is not possible using the standard \textit{Attention} formulation, as we will discuss in Section~\ref{shortcomings}. Although the promising results obtained by the authors, successive works \cite{DBLP:conf/sigir/BiAC20, DBLP:journals/corr/abs-2004-09424,DBLP:conf/www/JiangWRCYCGJC020} have shown that the \textit{Zero Attention Model} may perform inconsistently, and often it exposes equal or lower effectiveness than the standard \textit{Attention} formulation. 

In this paper, we first discuss some shortcomings of the standard \textit{Attention} formulation that prevent it from being an optimal solution when it comes to personalization. Then, we propose a novel \textit{Attention} variant, called \textit{Denoising Attention}, designed to solve these issues and show the benefits of our approach over other user modeling mechanisms.

\section{Preliminaries on Query-aware User Modeling}\label{preliminaries}

Users usually have diverse interests in multiple domains.
Although the representation of multifaceted user preferences is a powerful resource for personalization, not all those preferences are equally relevant to a specific user's information need.
For example, if a user is looking for a new book to read, her fashion-related preferences probably do not matter for personalizing the results of her current query.
\textit{Query-aware User Modeling} consists in building a user model at query time, based on previously gathered sources of user interest, by giving more importance to those related to the current search performed by the user.
In the literature, the definition of a user model with the previous characteristics has been provided by relying on the \textit{Attention} mechanism \cite{DBLP:journals/corr/BahdanauCB14}, which allows weighing the contribution of the user-related data w.r.t. the current search query.

In the following sections, we first describe the \textit{Attention} mechanism as it is usually employed in the context of \textit{Personalized Search}. Then, we discuss some shortcomings of its standard formulation when used for personalization.

\subsection{Attention Mechanism}\label{attention}

The \textit{Attention} mechanism, introduced by Bahdanau et al. \cite{DBLP:journals/corr/BahdanauCB14} in Neural Machine Translation, aims at computing a context vector by weighing the available contextual information w.r.t. a given input. A context vector can enrich the information carried by the input, helping, \textit{for example}, to disambiguate its meaning.

In \textit{Personalized Search}, the context vector is interpreted as the \textit{user's context vector}, \textit{i.e.}, the \textit{user model}; the contextual information is intended as the \textit{user's contextual information}, \textit{i.e.} the available user-related information, and the input is the \textit{search query}.
In the following, we assume that the user-related information sources and data are documents in the form of textual documents written by the users, previously accessed documents (\textit{e.g.}, web pages), user-generated content \cite{DBLP:journals/pervasive/KrummDN08} (\textit{e.g.}, product reviews or tweets), previously issued queries, or other content related to the users, and to their preferences.
At query time, the \textit{Attention} mechanism weighs the vector representations of these documents w.r.t. the query vector and aggregates them to produce the user model employed in the personalization process.
The \textit{Attention} mechanism comprises three steps aiming to build the context vector: \textit{scoring}, \textit{normalization}, and \textit{aggregation}.
The three steps are presented here below.

\paragraph{\textbf{Scoring}}\label{scoring}
First of all, an \textit{alignment model} \cite{DBLP:journals/corr/BahdanauCB14} (or \textit{scoring function}) $a$ is used to score how well the representations of the user-related documents match with the input query:
\begin{equation}\label{eq:score}
    e_{\bm{q}, \bm{d}} = a(\bm{q}, \bm{d})
\end{equation}
where, $\bm{d} \in \mathbb{R}^{m}$ and $\bm{q} \in \mathbb{R}^{n}$ are the vector representations of a user document and a given query, respectively, and $e_{\bm{q}, \bm{d}} \in \mathbb{R}$ is the matching score computed by the \textit{alignment model} $a: \mathbb{R}^{m} \times \mathbb{R}^{n} \to \mathbb{R}$ for the vectors $\bm{d}$ and $\bm{q}$.
It is important to outline that, usually, the dimension of the representations of the user document $\bm{d}$ and the query $\bm{q}$ are the same as they are projected in the same latent space. \\
The \textit{alignment model} can be as simple as the \textit{dot-product}, in which case we talk about \textit{Dot-Product Attention} \cite{DBLP:conf/emnlp/LuongPM15} and \textit{Scaled Dot-Product Attention} \cite{DBLP:conf/nips/VaswaniSPUJGKP17}, or the \textit{cosine similarity}, usually called \textit{Content-based Attention} \cite{DBLP:journals/corr/GravesWD14}.
Alternatively, the \textit{alignment model} can be 
a parameterized function, such as a Neural Network \cite{DBLP:journals/corr/BahdanauCB14}.

\paragraph{\textbf{Normalization}}\label{normalization}
The second step of the \textit{Attention} mechanism consists in the \textit{normalization} of the matching scores computed by the \textit{alignment model} to generate a probability distribution of the contextual information. The normalized matching scores are commonly called \textit{attention weights}.
This step is usually accomplished through the use of the \textit{Softmax} function \cite{DBLP:journals/corr/BahdanauCB14,DBLP:conf/emnlp/LuongPM15,DBLP:conf/nips/VaswaniSPUJGKP17,DBLP:journals/corr/GravesWD14}:
\begin{equation}\label{eq:norm}
    \alpha(\bm{q}, \bm{d}) = \softmax(e_{\bm{q}, \bm{d}}) = \frac{\exp(e_{\bm{q}, \bm{d}})}{\sum_{\bm{d^\prime} \in \bm{D_u}} \exp(e_{\bm{q}, \bm{d^\prime}})}
\end{equation}
where, 
$exp$ is the \textit{exponential} function, $\bm{D_u}$ is the set of all the documents related to the user $u$, and $\alpha(\bm{q}, \bm{d}) \in \mathbb{R}$ is the \textit{attention weight} of $\bm{d}$ w.r.t. $\bm{q}$.

\paragraph{\textbf{Aggregation}}\label{aggregation}
Finally, the third step consists in the \textit{aggregation} of the contextual information to produce the context vector $\bm{u}$, which, in our case, represents the \textit{user model}. This process is carried out by summing the user document vector representations weighed by their corresponding \textit{attention weights}:
\begin{equation}
    \bm{u} = \sum_{\bm{d} \in \bm{D_u}} \alpha(\bm{q}, \bm{d}) \cdot \bm{d}
\end{equation}

\subsection{Attention-based User Modeling Shortcomings}\label{shortcomings}

Although the \textit{Attention} mechanism allows building user models at query time, some shortcomings prevent it from being an optimal solution for personalization.
These issues are related to the \textit{Normalization} step and specifically to the use of the \textit{Softmax} and the \textit{exponential} function.
The \textit{Softmax} function, which is based on the \textit{Luce's choice axiom} \cite{Luce1959}, was proposed by Bridle \cite{DBLP:conf/nips/Bridle89} as a \textit{softened} (\textit{continuous} and \textit{differentiable}) generalization of the \textit{Arg\,max} function.
\textit{Arg\,max} is an operation that aims to find the point of the domain of a function in which it assumes maximum value.
\textit{Arg\,max} is neither continuous nor differentiable and, therefore, it does not allow for gradient-based optimizations \cite{Goodfellow-et-al-2016} (\textit{i.e.}, it cannot be used in Neural Networks, as it would break back-propagation).
\textit{Softmax} is primarily used by classifiers for computing a probability distribution over $n$ output classes.
During training, the model \textit{should} learn to maximize the output of the correct class.

Because of the nature of the \textit{Softmax} function, whose goal is to select one among $n$ options, and its use of the \textit{exponential}, a \textit{Softmax}-based user modeling approach naturally tends to skew the user representation towards a single user document, the one that best \textit{aligns} with the query.
Such characteristics are usually not ideal for personalization as the other user documents could concur to a more informed and balanced representation of the user interests and preferences.
\textit{For example}, given the following vector of \textit{alignment scores} $[7.0, 3.0, 1.0, -2.0]$, by applying Eq. \ref{eq:norm} for normalization, we obtain the following \textit{attention weights} $[0.9796, 0.0179, 0.0024, 0.0001]$.
These weights cause the user model to be strongly biased toward the contextual information contained in a single user document.

A possible solution could be to constrain the \textit{alignment function}'s output so that the \textit{Normalization} step cannot produce an \textit{overly narrow} probability distribution of the contextual information.
However, if, \textit{for example}, we constrain the \textit{alignment scores} near zero by using the \textit{cosine similarity} as the \textit{alignment model}, the \textit{Softmax} normalization will \textit{overly smooth} the scores, thus causing that noisy information will be injected into the user model and have a strong influence on the current search.
\textit{For example}, given the following vector of \textit{alignment scores} $[0.7, 0.3, 0.1, -0.2]$, by applying Eq. \ref{eq:norm} for normalization we obtain the following \textit{attention weights} $[0.3809, 0.2553, 0.2090, 0.1548]$.
These weights highly reduce the initial \textit{alignment scores}' diversity, giving considerable importance to the less relevant user-related information and penalizing the most relevant one.
Moreover, the user information source whose \textit{alignment score} with the query is negative, indicating very low relatedness, gets a positive \textit{attention weight}.
In this case, we can say that \textit{Softmax} promotes the presence of \textit{potentially} noisy information in the user model instead of penalizing or \textit{filtering} it out.
Note that, to properly work with Softmax, which produces higher values for the higher alignment scores among its input, an alignment model must assign high alignment scores to the user documents appropriate for personalization and low alignment scores to those potentially harmful.

Lastly, as the \textit{Softmax} normalizes its input into a probability distribution, it follows that the \textit{attention weights} from Eq. \ref{eq:norm} are all positive and sum up to $1$ \cite{DBLP:conf/nips/ChorowskiBSCB15}.
Even if all the alignment scores were zero or negative, the \textit{attention weights} would \textit{all} be positive and sum to $1$.
\textit{For example}, given the following vector of \textit{alignment scores} $[0.0, 0.0, 0.0, 0.0]$, by applying Eq. \ref{eq:norm} for normalization we obtain the following \textit{attention weights} $[0.25, 0.25, 0.25, 0.25]$.
The same happens when all the \textit{alignment scores} are negative: $[-7.0, -3.0, -1.0, -2.0] \rightarrow [0.0016, 0.0899, 0.6641, 0.2443]$.
As there will always be at least one positive \textit{attention weight} and the sum of the \textit{attention weights} will always be $1$, the context vector cannot be filtered out.
In the context of personalization, this means that the user's context vector will never be zero, causing the personalization of search results to be performed even when no source of user-related information is in line with her current search.
In such cases, personalization could hurt the effectiveness of the search engine instead of improving it.

\section{Denoising Attention Mechanism}\label{denoising}

In this section, we present our proposal to address the shortcomings of the standard \textit{Attention} when used for personalizing search results.
As extensively discussed in Section~\ref{shortcomings}, the principal issues of the \textit{Attention} are related to its \textit{normalization} step, described in Section~\ref{normalization}, and specifically to the use of the \textit{Softmax} function to produce the \textit{attention weights}.
To counteract these issues, we need a mechanism able to avoid \textit{overly narrowing} or \textit{overly smoothing} the \textit{attention weights}, which can cause the model either to focus only on a single source of user-related information or to reduce the diversity of their estimated importance.
Moreover, this mechanism should finely filter out noisy contextual information, thus preventing it from flowing into the user model.
Finally, it should zero out the user's context vector when personalization could harm the retrieval process, \textit{i.e.}, when \textit{all} the user-related information is noisy or irrelevant with respect to the current search.
In this regard, we propose the \textit{Denoising Attention} mechanism.
The \textit{Denoising Attention} mechanism departs from the \textit{Softmax} function by adopting a more straightforward and robust normalization scheme, and it introduces a filtering mechanism based on the \textit{Rectifier Linear Unit} \cite{DBLP:conf/icml/NairH10} and a threshold value.
To complement those changes, we rely on a \textit{cosine similarity}-based alignment model to evaluate the relatedness of the sources of user-related information w.r.t. the current search.

\paragraph{\textbf{Scoring}}\label{denoising_scoring}

For an \textit{alignment model} to act in a complementary way with the changes introduced in the next paragraph, we need a function $a(\bm{q}, \bm{d})$ bounded between 0 and 1, as an unbounded function would make it difficult to control which information flows into the user model.
To compute the \textit{alignment scores} $e_{\bm{q}, \bm{d}}$, we then rely on the following \textit{cosine similarity}-based function: 
\begin{equation}\label{eq:denoising_alignment}
    e_{\bm{q}, \bm{d}} = a(\bm{q}, \bm{d}) = \frac{\cos(\bm{q}, \bm{d}) + 1}{2}
\end{equation}
This function shifts the cosine similarity codomain to $[0, 1]$.

\paragraph{\textbf{Filtering}}\label{denoising_filtering}
The first change we propose to the standard \textit{Attention} mechanism is the explicit addition of a \textit{filtering} step.
First, we introduce a \textit{threshold} parameter $t$ that we use to \textit{negativize} the alignment scores of the user data
loosely related to the input query.
In Section \ref{sub:model_analysis} we show in detail the impact it has on the model's performance.
We call this operation \textit{alignment scores shifting} and we define it as follows:
\begin{equation}\label{eq:threshold}
    \mathit{shifted\_e}_{\bm{q}, \bm{d}} = e_{\bm{q}, \bm{d}} - \sigma(t)
\end{equation}
where $\sigma$ is a \textit{squashing function} \cite{DBLP:journals/nn/HornikSW89}, which allows us to constrain $t$ in $[0, 1]$ during training. We relied on the \textit{Sigmoid} function in our final implementation.
Secondly, we apply the \textit{Rectifier Linear Unit} (\textit{ReLU}) \cite{DBLP:conf/icml/NairH10} to the \textit{shifted alignment scores}. 
\textit{ReLU} is a very popular activation function used in Neural Networks, and it is formulated as follows:
\begin{equation}\label{eq:relu}
    \relu(x) = \max\{0, x\}
\end{equation}
What makes \textit{ReLU} convenient in the personalization context is its ability to zero out negative values, in our case, the \textit{shifted alignment scores} of noisy user-related information, while leaving unaltered the others:
\begin{equation}\label{eq:filter}
    \mathit{filtered\_e}_{\bm{q}, \bm{d}} = \relu(\mathit{shifted\_e}_{\bm{q}, \bm{d}})
\end{equation}
By combining these two operations, we can both control the information flow from the user data to the user model in diverse search scenarios and filter out the noisy user-related information that could harm the retrieval process.
To avoid the well-known dying ReLU problem \cite{DBLP:journals/corr/abs-1903-06733,DBLP:journals/corr/abs-1803-08375} and let the model learn to zero out the user model correctly, we sum the user model to the query representation during training.

\paragraph{\textbf{Normalization}}
The second major change we propose to the standard \textit{Attention} mechanism is the use of the \textit{plain} normalization operation in place of the \textit{Softmax} for the computation of the \textit{attention weights}, which is defined as follows:
\begin{equation}\label{eq:plain_norm}
    \alpha(\bm{q}, \bm{d}) = \frac{\mathit{filtered\_e}_{\bm{q}, \bm{d}}}{\sum_{\bm{d^\prime} \in \bm{D_u}} \mathit{filtered\_e}_{\bm{q}, \bm{d^\prime}}}
\end{equation}
As the \textit{plain} normalization can cause numerical instability when all the $\mathit{filtered\_e}_{\bm{q}, \bm{d}}$ are zero, we slightly change Eq.~\ref{eq:plain_norm} into:
\begin{equation}\label{eq:denoising_norm}
    \alpha(\bm{q}, \bm{d}) = \frac{\mathit{filtered\_e}_{\bm{q}, \bm{d}}}{
    \max \left\{ \sum_{\bm{d^\prime} \in \bm{D_u}} \mathit{filtered\_e}_{\bm{q}, \bm{d^\prime}}, \varepsilon \right\}}
\end{equation}
where $\varepsilon$ is a very low positive value.

Note that the proposed filtering mechanism cannot work correctly with \textit{Softmax}.
First of all, the \textit{Softmax} function is \textit{translationally invariant}, which means that adding or removing the same value to all the components of its input does not change its output.
For example, $\softmax([0.7, 0.3, 0.1, -0.2]) = \softmax([0.7, 0.3, 0.1, -0.2]$ $- 0.3) = [0.3809, 0.2553, 0.2090, 0.1548]$.
Secondly, zeroing out negative values through \textit{ReLU} does not prevent the \textit{Softmax} from producing positive attention weights for those values, as already shown in Section~\ref{shortcomings}.
On the contrary, Eq.~\ref{eq:denoising_norm} does not suffer from those issues and can produce zero \textit{attention weights}.
By avoiding the use of the \textit{Softmax} and its \textit{exponential} function, the \textit{Denoising Attention} \textit{normalization} step does not suffer from many of the \textit{Attention} shortcomings discussed in Section~\ref{shortcomings}.

\paragraph{\textbf{Aggregation}}
The aggregation of the user-related information is performed as a linear combination, following the standard \textit{Attention} formulation (Section~\ref{aggregation}).
However, as normalizing with Eq.~\ref{eq:denoising_norm} allows to produce zero \textit{attention weights}, the \textit{aggregation} step can yield a \textit{zero context vector}, which ultimately allows avoiding personalization when no source of user information is related to her current search.

\paragraph{\textbf{Denoising Attention Weights}}
To sum up, we propose to compute the weights for the user-related information as follows:
\begin{equation}\label{eq:denoising}
    \alpha(\bm{q}, \bm{d}) = \frac{\relu\left(e_{\bm{q}, \bm{d}} - \sigma(t)\right)}{\sum_{\bm{d^\prime} \in \bm{D_u}} \relu\left(e_{\bm{q}, \bm{d^\prime}} - \sigma(t)\right)}
\end{equation}
In contrast with the standard \textit{Attention} formulation, \textit{Denoising Attention} is able to 1) selectively filter out the noisy contextual information from the user-related data
before aggregating them in the context vector, and 2) zero out the context vector when all the sources of user-related information are unrelated to her current search.
Moreover, the combined use of our \textit{filtering mechanism} and \textit{normalization} function makes our \textit{Attention} variant prone to avoid \textit{overly narrow} or \textit{overly smooth} attention weights.
This way, the model preserves the estimated importance of the user-related information sources and does not focus only on one of them, thus composing a balanced representation of the user preferences related to the current query while filtering those unrelated.
For a sake of comparison, the \textit{alignment scores} $[0.7, 0.3, 0.1, -0.2]$ produce the \textit{attention weights} $[0.3809, 0.2553, 0.2090, 0.1548]$ when fed to Eq. \ref{eq:norm}, whereas they produce the \textit{attention weights} $[0.75, 0.25, 0.0, 0.0]$ when fed to Eq. \ref{eq:denoising} with $\sigma(t) = 0.1$.

\section{Personalized Results Re-Ranking}\label{re-ranking}

In this section, we approach the problem of exploiting the user model for personalization and we introduce the task we have considered for evaluating the proposed user modeling approach, i.e. \textit{Personalized Results Re-Ranking}.
Moreover, we describe the personalized re-ranking framework we employed for the comparative evaluation reported in the following sections. This framework allowed us to test different user modeling techniques with ease.

\begin{figure}[ht]
  \centering
  \includegraphics[width=1.0\textwidth]{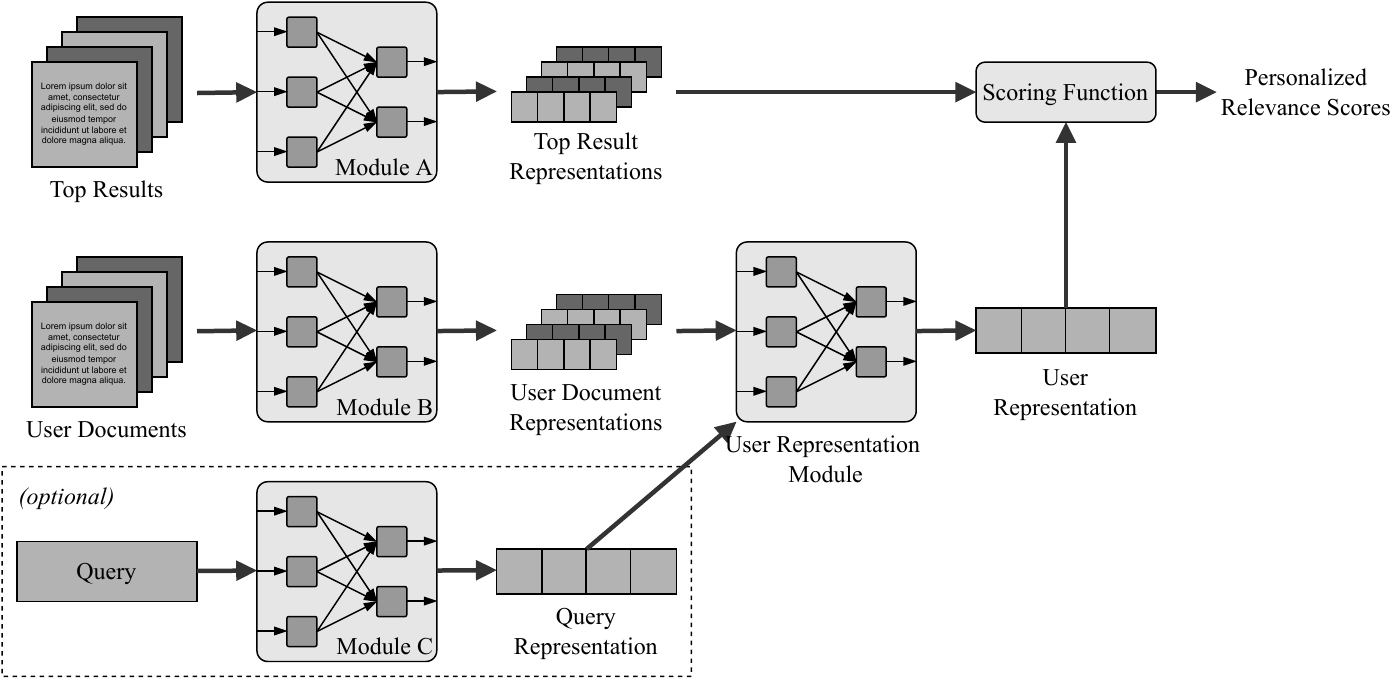}
  \caption{Personalized Results Re-Ranking Framework.}
  \label{fig:framework}
\end{figure}

In \textit{Personalized Results Re-Ranking}, a retrieval system (first stage retriever) computes a ranked list of documents in response to a search query.
Then, a personalization component computes new relevance scores for the initially retrieved documents by leveraging the user-related information.
Finally, the personalized relevance scores, or a combination of those and the scores computed by the first stage retriever, are used to re-rank the initially retrieved list of documents.
In the latter case, the two relevance scores are usually aggregated via convex combination:
\begin{equation}\label{eq:score_fusion}
    \mathit{final\_score} = ( 1 - \lambda ) \cdot a + \lambda \cdot b
\end{equation}
where, $a$ and $b$ are the relevance scores computed by the first stage retriever and the personalization model, respectively, and $\lambda$ is a parameter that controls the influence of the two on the final score.

Fig. \ref{fig:framework} depicts the \textit{Personalized Results Re-Ranking Framework} we relied on for comparing various user modeling techniques for \textit{Personalized Search} (Section~\ref{experimental_setup} and \ref{eval}).
The framework comprises two modules that generate the vector representations of the top-$k$ results retrieved by the first stage retriever and those of the user-related documents. Once computed the user-related document representations, the \textit{user representation module} aggregates them into the user model.
In the case of query-aware user modeling, as the query is involved in weighing the contribution of each user-related document, an additional module is employed to produce the query representation used in that process.
Finally, a \textit{scoring function} computes a personalized relevance score for each initially retrieved result by comparing its representation with the user representation.
Those scores are then combined with the first stage retriever's scores as in Eq. \ref{eq:score_fusion}.

In the experiments presented in Section~\ref{eval}, we relied on TinyBERT \cite{DBLP:conf/emnlp/JiaoYSJCL0L20} followed by a \textit{mean pooling} operation to embed both the top retrieved documents, the user information, and the query (\textit{if needed}), as they all are in text format in the dataset we employed for the evaluation.
TinyBERT is a distilled \cite{DBLP:journals/corr/HintonVD15,DBLP:journals/ijcv/GouYMT21} version of the well-known Transformer-based \cite{DBLP:conf/nips/VaswaniSPUJGKP17} Neural Language Model BERT \cite{DBLP:conf/naacl/DevlinCLT19}.
We chose TinyBERT due to its training and inference speed, lightweight GPU memory consumption, and, above all, due to our hardware limitations (for experimentation, we used an NVidia\textsuperscript{\textregistered} RTX 2080 Ti GPU with 11 GBs of VRAM, which is not enough to fine-tune BERT in our specific setting).
As with many recent ranking models \cite{DBLP:conf/emnlp/ReimersG19,DBLP:conf/ecir/GaoDCFDC21,DBLP:conf/emnlp/KarpukhinOMLWEC20,DBLP:conf/acl/LeeCT19,DBLP:conf/sigir/MacAvaneyN0TGF20b} that rely on learned dense representations of documents \cite{DBLP:series/synthesis/2021LinNY}, we employed the \textit{cosine similarity} as our \textit{scoring function}.

As the main purpose of the contribution we present in this paper is to propose a novel Attention variant for defining user models at query time, the simple technique we implement for re-ranking aims at making it possible to comparatively evaluate the effectiveness of the proposed model with alternatives at the state-of-the-art with ease, allowing us to switch the compared user models seamlessly.
We also point out that the authors of previous contributions did not share their code or did not provide adequate instructions to reproduce the results they obtained or train/run their proposed models, posing several reproducibility issues.

\section{Experimental Setup}\label{experimental_setup}

The experiments reported in this section aim to answer the following research questions:

\begin{research_questions}
    \item \label{rq1} Do the query-aware \textit{Attention}-based user models increase the effectiveness of a personalized retrieval model w.r.t. simpler operations for aggregating user data?

    \item \label{rq2} Does the \textit{Denoising Attention}-based user model increase the effectiveness of a personalized retrieval model w.r.t. other \textit{Attention}-based user models?
    
    \item \label{rq3} Is the query-aware user representation produced by the \textit{Denoising Attention} better-balanced w.r.t. the query-related user preferences than those of other \textit{Attention} variants?
    
    \item \label{rq4} Is the \textit{Denoising Attention}-based personalization more \textit{robust}, \textit{i.e.}, less likely to decrease the system’s effectiveness due to noisy user-related data, than the other considered approaches?
\end{research_questions}
To answer the research questions \ref{rq1} and \ref{rq2}, we conducted a comparative evaluation of the retrieval effectiveness of the personalized re-ranking pipeline described in Section~\ref{re-ranking} using several different user models.
Then, we compared the retrieval effectiveness of the user models for the queries personalized by the \textit{Denoising Attention}-based model to answer question \ref{rq3}.
Finally, to answer the research question \ref{rq4}, we compared the number of times the considered user models decreased the retrieval effectiveness of our first-stage retriever, BM25.

In the following, we present the datasets we employed for conducting our evaluations (Section~\ref{dataset}), we introduce the baselines we have selected (Section~\ref{baselines}), and we outline the training setup and evaluation procedure (Section~\ref{setup}).
We make all our code available for future works and reproducibility purposes\footnote{We will add a link to the repository upon acceptance}.

\subsection{Datasets}\label{dataset}

To conduct our experimental evaluations, we relied on two datasets that account for different search scenarios.
First, we considered a Web Search dataset based on the AOL query log \cite{DBLP:conf/infoscale/PassCT06}.
Then, we relied on a synthetic dataset we built following the procedure described by Tabrizi et al. \cite{DBLP:journals/ipm/TabriziSZT18} to simulate a domain-specific search scenario, in our case Academic Search.
We describe both datasets in detail in the following sections.

\begin{table}[t]
    \caption{Statistics of the employed datasets.}
    \centering
    
    \begin{tabular}{lc|lc}
    \toprule
    \multicolumn{4}{c}{\textbf{Web Search Dataset}} \\
    \midrule
    \# documents & $1\,291\,695$     & \# users & $30\,166$ \\
    \# train queries & $212\,386$    & avg. query length & $3.57 \pm 1.51$ \\
    \# val queries & $31\,064$ & avg. relevants & $1.15 \pm 0.46$ \\
    \# test queries & $36\,052$      & avg. user docs & $136.62 \pm 134.17$  \\
    \midrule
    \multicolumn{4}{c}{\textbf{Academic Search Dataset}} \\
    \midrule
    \# documents & $4\,201\,265$     & \# users & $63\,738$ \\
    \# train queries & $419\,004$    & avg. query length & $7.53 \pm 2.64$ \\
    \# validation queries & $4\,241$ & avg. relevants & $5.33 \pm 5.11$ \\
    \# test queries & $24\,056$      & avg. user docs & $53.59 \pm 50.94$  \\
    \bottomrule
    \end{tabular}
    
    \label{tab:datasets}
\end{table}

\subsubsection{Web Search Dataset}

Thanks to its potential, personalization in Web Search has been a hot topic for many years and has attracted the attention of several researchers both from academia (\cite{DBLP:conf/www/HannakSKKLMW13,DBLP:conf/cikm/SiegMB07,DBLP:conf/semweb/NollM07}) and private companies (\cite{DBLP:conf/sigir/TeevanDH05,DBLP:conf/sigir/TeevanDL08,DBLP:journals/tochi/TeevanDH10}).
Users search for a myriad of information on the Web, and discerning among the diverse - and often \textit{unrelated} - interests a user can have could highly influence the personalization effectiveness in this scenario.
This characteristic makes Web Search a perfect fit to evaluate the \textit{Attention} variant we propose in this work.

The AOL query log \cite{DBLP:conf/infoscale/PassCT06} is one of the most known large-scale set of data for the evaluation of \textit{session-based} personalization models (\cite{DBLP:conf/iclr/AhmadCW18,DBLP:conf/sigir/AhmadCW19,DBLP:conf/sigir/YaoDW20,DBLP:conf/sigir/ZhouDW20,DBLP:conf/sigir/LuDXWW20,DBLP:conf/cikm/ZhouD0W21,DBLP:conf/www/YaoDXW20,DBLP:journals/tois/YaoDW22,DBLP:conf/wsdm/Deng0D22}).
Although we are not focusing on \textit{session-based} personalization, we can rely on this same query log to evaluate the effectiveness of our proposal.
Unfortunately, the authors of the previous works that relied on the AOL query log did not release the instructions to re-build the datasets they employed in their evaluations.
Therefore, we derive a novel Web Search dataset suited for our evaluation from it.
We acknowledge the availability of other large-scale Web Search datasets, such as the Yandex\footnote{\url{https://www.kaggle.com/competitions/yandex-personalized-web-search-challenge}} dataset, but, unfortunately, those datasets provide only anonymized texts for queries and documents, they are not publicly available, or they lack user unique identifiers.

We now review the procedure we followed to build our Web Search dataset from the AOL query log and make all the scripts to re-build such a dataset available for future research\footnote{We will add a link to the repository upon acceptance}.

\paragraph{\textbf{Retrieving documents}}

A noticeable limitation of the AOL query log is that it does not provide the document contents but only the URL of clicked documents (\textit{if any}).
Because the logs date back to 2006, many of the clicked URLs are not available today, or the content of the documents they point to has changed since users accessed them.
To retrieve document contents similar to those seen by the users when the logs were collected, we relied on the recently proposed \textit{aolia-tools} \cite{DBLP:conf/ecir/MacAvaneyMO22}, which leverage the Internet Archive's Wayback Machine service.
We refer the reader to MacAvaney et al. \cite{DBLP:conf/ecir/MacAvaneyMO22} for an in-depth discussion on this topic.
Once retrieved the document contents, we identified and removed non-English documents by analyzing them using Google's \textit{Compact Language Detector} v3\footnote{\url{https://github.com/google/cld3}}.

\paragraph{\textbf{Query logs cleaning}}

The AOL query log comprises queries issued by real users between March 1, 2006, and May 31, 2006.
To derive a dataset suited for our evaluation, we operated a cleaning process aiming at obtaining an high quality query set while reducing noise that could interfere in our evaluation.
First, we discarded all the queries with no related clicks from the query log and those pointing to non-English documents.
Then, following MacAvaney et al. \cite{DBLP:conf/ecir/MacAvaneyMO22}, which reported that several queries from the AOL query log have a navigational nature, we discarded those containing Internet domain references (e.g., \textit{.com}, \textit{.org}, etc.) or website names as they do not need personalization and can be easily identified during pre-processing.
For ethical reasons, we also discarded all the queries containing or pointing to adult or illegal contents.
Following Sordoni et al.~\cite{DBLP:conf/cikm/SordoniBVLSN15}, we removed non-alphanumeric characters from the queries, applied a spelling corrector (\textit{SymSpell}\footnote{https://github.com/wolfgarbe/SymSpell}) and lower-cased the queries. Then, we discarded all the queries shorter than three characters.
To avoid introducing in the test set \textit{$\langle$query, user, document$\rangle$} triplets also present in the train set, we kept only the first appearance of such triplets by comparing their associated timestamps.
Note that we are not interested in re-finding behavior in our work \cite{DBLP:conf/cikm/TylerWZ10}.

\paragraph{\textbf{Training / Validation / Test Splits}}
Following previous works \cite{DBLP:conf/cikm/SordoniBVLSN15,DBLP:conf/sigir/AhmadCW19}, we considered the queries formulated in the first five weeks as a background set. We discarded all the queries from users with less than $20$ associated queries in this set to ensure having enough user-related data to conduct personalization, which, in this case, is based on previously accessed web pages.
We then temporally split the remaining weeks' worth of queries.
We used six weeks for \textit{training} queries, one week for \textit{validation} queries, and one week for \textit{test} queries.
Then, we fine-tuned the hyper-parameters of BM25 \cite{DBLP:conf/sigir/RobertsonW94}, which we use as our first stage retriever in the experimental evaluation, on thousands of non-test queries.
Finally, we discarded the queries for which BM25 does not retrieve any relevant document in the top 1000 results. Likewise, for the remaining ones, we retain only the relevant documents present in the top 1000 results retrieved by BM25, as we are interested in results re-ranking.
\\
Table \ref{tab:datasets} reports the statistics of the final dataset.

\subsubsection{Academic Search Dataset}

Alongside Web Search, Domain-specific Search is a popular research topic nowadays.
Unlike Web Search, in domain-specific search scenarios, the user interests are more focused on particular topics, which could make finely discerning among the user-related data to pick those most promising for personalizing the current search conducted by the user more challenging.

Due to the lack of a publicly available Domain-specific Search dataset for studying personalization, researchers have recently tackled personalization in Product Search scenarios relying on synthetic datasets built upon product reviews from a popular e-commerce platform \cite{DBLP:conf/sigir/AiZBCC17,DBLP:journals/corr/abs-2004-09424,DBLP:conf/sigir/BiAC20,DBLP:conf/sigir/BiAC21,DBLP:journals/inffus/BassaniP22}.
However, due to the number of different queries present in these datasets, a few hundred in most cases, and their low quality \cite{DBLP:journals/inffus/BassaniP22}, we did not employ them in our comparative evaluation.
Instead, we followed the procedure described by Tabrizi et al. \cite{DBLP:journals/ipm/TabriziSZT18} to build an Academic Search dataset that allow us to test our \textit{Attention} variant in a domain-specific search scenario.
In particular, we relied on the \textit{ArnetMiner}'s \textit{Citation Network Dataset V12} \cite{DBLP:conf/kdd/TangZYLZS08}, which makes available the metadata, such as \textit{titles}, \textit{abstracts}, list of \textit{authors}, and list of \textit{citations}, of  $4\,894\,081$ academic papers (papers' full texts are not available).
We now describe the process we followed to build our dataset.

\paragraph{\textbf{Query Generation}}

Firstly, we removed all non-English papers by leveraging Google's \textit{Compact Language Detector} v3\footnote{\url{https://github.com/google/cld3}}.
Then, following the approach described by Tabrizi et al. \cite{DBLP:journals/ipm/TabriziSZT18}, we generated user-query-document triplets as follows: for each academic paper, we considered its title as a query, the list of its citations as the documents relevant to that query, and we assumed that the first author is the user submitting the query.
Tabrizi et al.~\cite{DBLP:journals/ipm/TabriziSZT18} proposed other methods to generate synthetic queries from research papers, but they only reported the evaluation of the one we employ here.
As the titles of academic papers are written in well-formed natural language, we applied stop-word removal using the \textit{NLTK}'s \cite{DBLP:books/daglib/0022921} stop-words list and a non-destructive stemmer, \textit{i.e.}, the \textit{Krovetz stemmer} \cite{DBLP:conf/sigir/Krovetz93}, to obtain queries that resemble real-world ones.
Finally, we discarded all the generated queries whose related users have less than 20 associated documents, \textit{i.e.}, published papers.

\paragraph{\textbf{Training / Validation / Test Splits}}

We split the obtained dataset into \textit{training} and \textit{test} sets chronologically, \textit{i.e.} by using the queries generated from papers published \textit{after} 2018 as the \textit{test set}. We then randomly split the \textit{training set} to obtain a \textit{training set} and a \textit{validation set}, using a splitting ratio of $99:1$.
We opted for a chronological \textit{training} / \textit{test} split instead of a random partitioning so that the dataset is closer to a real scenario, where all the searches in the \textit{test set} happen after the searches in the \textit{training set}.

\paragraph{\textbf{Dataset Refining}}
As discussed by Tabrizi et al.~\cite{DBLP:journals/ipm/TabriziSZT18}, not all the references of a paper are necessarily relevant (from an Information Retrieval perspective) to the topic expressed by its title, which we use as a query.
The authors, however, claim that since mistakenly considering some irrelevant documents as relevant will be the same for all the compared models, their presence does not violate the fairness of the comparisons if evaluation measures are averaged over many queries (\textit{thousands}, in our case).
To reduce the presence of spurious relevant documents and malformed queries, we applied simple heuristics, similarly to Tabrizi et al~\cite{DBLP:journals/ipm/TabriziSZT18}.
As we are comparing different user modeling techniques in the context of Personalized Results Re-Ranking, we consider well-formed only the queries for which BM25 \cite{DBLP:conf/sigir/RobertsonW94}, which we use as a first-stage retriever, retrieves relevant documents in the top results.
Although this is a strong assumption, the reader should consider two key factors.
First, every re-ranking approach is inherently bounded by the recall of the first stage retriever.
Therefore, as all the compared models re-rank the BM25's results, they all share the same limitations.
Secondly, only an exiguous subset of a paper's references can be considered relevant w.r.t. its main topic.
For example, if considering our paper, we would consider only the works concerning query-aware personalization, presented in Section~\ref{related_works}, as truly relevant documents for a query built using our title.
Although filtering the references by their positioning in the paper (the section in which they appear) \textit{could be} a better solution in this case, we lack such information.
From the original query set, we removed all the queries for which BM25 does not retrieve any relevant document in the top-1000 results.
Likewise, for each of the remaining queries, we retain only the relevant documents present in the top 1000 results retrieved by BM25.
Before computing the BM25 results, to maintain most queries and relevant documents as possible, we fine-tuned its parameters on the validation queries.
Table \ref{tab:datasets} reports the statistics of the final dataset.
We make the dataset and all the scripts to re-build it available for future research\footnote{We will add a link to the repository upon acceptance}.

\subsection{Baselines}\label{baselines}

In this section, we introduce the baselines employed in our comparative evaluation.
We compared the \textit{Denoising Attention}-based user model with user models based on the standard \textit{Attention} formulation, the \textit{Zero Attention strategy} proposed by Ai et al.~\cite{DBLP:conf/cikm/AiHVC19}, and the \textit{Multi-Head Attention} \cite{DBLP:conf/nips/VaswaniSPUJGKP17}.
We also considered a user model built by simply averaging the user-related documents’ representations to assess whether \textit{Attention}-based user models can improve over simpler operations for aggregating user data.
%
For reference, we also performed comparison with the classic probabilistic retrieval model BM25 \cite{DBLP:conf/sigir/RobertsonW94}, which we used as first stage retriever.

\begin{itemize}
    \item \textbf{Attention:}
    The first baseline is a query-aware user model based on the standard \textit{Attention} formulation.
    
    \item \textbf{Zero Attention:}
    The second baseline is a query-aware user model based on the \textit{Zero Attention Strategy} proposed by Ai et al. \cite{DBLP:conf/cikm/AiHVC19}. The \textit{Zero Attention Strategy}, introduced in Section~\ref{related_works}, was proposed to automatically determine \textit{when} and \textit{how} to conduct personalization.
    
    \item \textbf{Multi-Head Attention:}
    The third baseline is a query-aware user model based on the \textit{Multi-Head Attention} \cite{DBLP:conf/nips/VaswaniSPUJGKP17}, a scaled-up variant of the \textit{Attention} mechanism. \textit{Multi-Head Attention allows the model to jointly attend to information from different representation subspaces} \cite{DBLP:conf/nips/VaswaniSPUJGKP17}.
    We use 4 \textit{Attention} heads in our experiments.
    
    \item \textbf{Mean:}
    The fourth baseline is \textit{static} user model that computes user representations as the arithmetic mean of the user-related documents' representations.
    As averaging is a simple form of vector aggregation, its addition to the evaluation allows us to assess whether query-aware user modeling techniques are \textit{truly} beneficial.
    
\end{itemize}
We trained three variants for both the \textit{Attention}-based and the \textit{Zero Attention}-based user models by employing different \textit{alignment functions}.
The first variant employs the \textit{scaled-dot product}, popularized by the Transformer architecture \cite{DBLP:conf/nips/VaswaniSPUJGKP17}.
The second one uses the \textit{cosine similarity}, similarly to our \textit{Denoising Attention}.
The latter relies on the \textit{alignment model} proposed by Bahdanau et al. \cite{DBLP:journals/corr/BahdanauCB14} in the paper where the \textit{Attention} mechanism was proposed, which is commonly called \textit{Additive Attention} \cite{DBLP:conf/nips/VaswaniSPUJGKP17}.

We leave experimentation and comparison with Transformer models \cite{DBLP:conf/nips/VaswaniSPUJGKP17} for future work. 
We note that our proposed \textit{Attention} variant could also be used in place of the standard formulation in these complex architectures as well.

\subsection{Setup \& Evaluation Metrics}\label{setup}
We relied on ElasticSearch \cite{gormley2015elasticsearch} for BM25, HuggingFace's Transformers \cite{DBLP:journals/corr/abs-1910-03771} for TinyBERT\footnote{\url{https://huggingface.co/huawei-noah/TinyBERT_General_4L_312D}}, and PyTorch \cite{DBLP:conf/nips/PaszkeGMLBCKLGA19} for the implementation of all the neural models.
We fine-tuned BM25's $k1$ and $b$ parameters on thousands of non-test data before computing the result lists for all the queries.
BM25 scores were computed on the concatenation of documents' title and body (\textit{the papers' abstracts}) by first removing stop-words and applying the Krovetz stemmer \cite{DBLP:conf/sigir/Krovetz93}.
We trained each variation of the \textit{Personalized Results Re-Ranking Framework} introduced in Section~\ref{re-ranking}
on an NVidia\textsuperscript{\textregistered} RTX 2080 Ti GPU for 20 epochs using a hinge loss \cite{DBLP:conf/esann/WestonW99} defined over a triplet, similarly to \cite{DBLP:conf/ecir/GaoDCFDC21}, with a margin set to $0.1$ (\textit{we found larger margins to decrease performances in all cases}), and AdamW optimizer \cite{DBLP:journals/corr/abs-1711-05101,DBLP:conf/iclr/LoshchilovH19} with learning rate set to $5\times10^{-5}$, and batch size set to 32.
We train the model with hard negatives sampled from the top results retrieved by BM25 and in-batch random negatives samples.
During training, due to our hardware limitations, we randomly sampled the titles of $20$ user documents to use for personalization, while during the evaluation, we used those from all the available user documents.
After training, we fine-tuned the $\lambda$ parameter of Eq.~\ref{eq:score_fusion} and the \textit{Denoising Attention}'s threshold with grid-search on the \textit{validation set}.
In the case of the Academic Search Dataset, during the experiments, for each query, we skipped all the documents published after the release of the manuscript used for generating the query as none of them has relevant judgments for it because of dataset construction and, therefore, they would only add noise to the evaluation as suggested by Tabrizi et al. \cite{DBLP:journals/ipm/TabriziSZT18}.
For the final evaluation, we re-ranked the top 1000 results retrieved by BM25 with each of the considered user models.

To evaluate the effectiveness of the compared models, we employed 1) \textit{Mean Average Precision} (MAP), 2) \textit{Mean Reciprocal Rank} (MRR), and 3) \textit{Normalized Discounted Cumulative Gain} (NDCG).
MRR and NDCG were computed on the top 10 documents retrieved by each model, whereas MAP was computed on the top 100.
Statistical significance testing was conducted using a Bonferroni corrected Fisher's randomization test \cite{DBLP:conf/cikm/SmuckerAC07} with $p < 0.001$.
Metrics computation and comparison were conducted using the Python evaluation library \texttt{ranx} \cite{DBLP:conf/ecir/Bassani22}.

\section{Results and Discussion}\label{eval}

In this section, we present the results of our comparative evaluations.
First, we discuss the retrieval effectiveness of the personalized re-ranking pipeline described in Section~\ref{re-ranking} when considering different user modeling techniques.
Second, we evaluate how balanced the user preferences expressed by the considered user models are w.r.t. each query.
Third, we analyze the robustness of the compared user models, evaluating the probability they decrease the system’s effectiveness in the presence of noisy user-related data.
Finally, we analyze the performance of our proposed \textit{Attention} variant from multiple perspectives and ablate our proposal.
We remind the reader that the only difference between the compared personalization models is the technique used for defining the user model, while the other components are the same for all the considered models.

\subsection{Overall Retrieval Effectiveness}\label{overall-effectiveness}

In this section, we review the results obtained when re-ranking the documents by combining the relevance score produced by different user models with the document's relevance score produced by BM25, as described in Section~\ref{re-ranking}, aiming to answer questions \ref{rq1} and \ref{rq2}.

As reported in Table \ref{tab:framework_results}, combining personalized relevance scores with those coming from our first stage retriever, BM25, improved the retrieval effectiveness of the latter regardless of the user modeling mechanism employed, thus confirming the utility of personalization in both the considered search scenarios and datasets.

The \textit{Attention} and \textit{Zero Attention}-based user models generally improved over the \textit{Mean} user model.
However, we notice this is not always the case.
Of all the different variants, only those using the \textit{scaled-dot} as an \textit{alignment model} significantly improved over \textit{Mean} on both the considered datasets.
On the other hand, those relying on the \textit{additive} and the \textit{cosine} \textit{alignment models} achieved mixed results, sometimes even decreasing w.r.t. \textit{Mean}.
Moreover, we highlight that in the case of the \textit{Web Search Dataset}, the best-performing \textit{Attention} baselines' improvements are not that pronounced (only 3\% in MAP, MRR, and NDCG, respectively).
The \textit{Zero Attention}-based user models generally achieved slightly worse results than their \textit{Attention}-based counterparts, which raises questions regarding the efficacy of its employed mechanism for conducting differentiated personalization.
Our findings on the \textit{Zero Attention}-based user model are consistent with results from previous works \cite{DBLP:conf/sigir/BiAC20,DBLP:journals/corr/abs-2004-09424,DBLP:conf/www/JiangWRCYCGJC020}.
The results obtained by both the standard \textit{Attention}-based user model and the \textit{Zero Attention}-based user model with the \textit{Cosine Similarity} as \textit{alignment model} confirm that constraining the \textit{alignment scores} causes noisy information to leak in the user model, as discussed in Section~\ref{shortcomings}.
Finally, the \textit{Multi-Head Attention}-based user model's results are among the lowest for both datasets.
The additional complexity introduced by this approach did not deliver improvements over the other \textit{Attention}-based models while introducing additional overhead.
We suspect this is because we are employing high-quality text representations obtained using TinyBERT \cite{DBLP:conf/emnlp/JiaoYSJCL0L20}, a distilled version of the well-known Transformer \cite{DBLP:conf/nips/VaswaniSPUJGKP17} model BERT \cite{DBLP:conf/naacl/DevlinCLT19}, which alleviates the need for \textit{attending information from different representation sub-spaces} \cite{DBLP:conf/nips/VaswaniSPUJGKP17} during personalization.
We leave further investigation for future work.
If we consider only the \textit{Attention}-based user model with the \textit{scaled-dot} \textit{alignment model}, the obtained results positively answer our first research question, \ref{rq1}.
However, this is not the case for all the other \textit{Attention} baselines, which confirms the need for the in-depth investigation we are conducting on the use of the \textit{Attention} mechanism for query-aware personalization.

When employing the \textit{Denoising Attention}-based user model, the results re-ranking pipeline achieved substantial improvements over both that using the \textit{Mean}-based user model and those relying on \textit{Attention}-based user models, corroborating our intuitions about the shortcomings of the standard \textit{Attention} formulation when it comes to personalization (Section~\ref{shortcomings}) and the advantages brought by our proposal (Section~\ref{denoising}).
In particular, it improved over \textit{Mean} by 20\%, 22\%, and 19\% in MAP, MRR, and NDCG, respectively, on the \textit{Web Search Dataset} and by 23\%, 15\%, and 21\% in MAP, MRR, and NDCG, respectively, on the \textit{Academic Search Dataset}.
Moreover, it increased over the best-performing \textit{Attention}-based baseline by 17\%, 18\%, and 16\% in MAP, MRR, and NDCG, respectively, on the \textit{Web Search Dataset} and by 14\%, 10\%, 13\% in MAP, MRR, and NDCG, respectively, on the \textit{Academic Search Dataset}.
Finally, it enhanced the BM25 effectiveness by 38\%, 41\%, and 40\% in MAP, MRR, and NDCG, respectively, on the \textit{Web Search Dataset} and by 50\%, 29\%, and 41\% in MAP, MRR, and NDCG, respectively, on the \textit{Academic Search Dataset}.
The obtained results clearly show the robustness of our proposed \textit{Attention} variant to search scenarios with noticeable structural differences.
On one side, for \textit{Web Search}, it is fundamental to finely select among the user-related data those most promising for conducting personalization to improving over standard operations for building user models, \textit{i.e.}, averaging over the representations of the user-related data.
On the other side, in the case of \textit{Academic Search}, user-related information is very focused and, therefore, it is easier to improve a user model that averages the representations of the user-related data by simply weighing the contribution of those data w.r.t. the current query.
Nonetheless, \textit{Denoising Attention} still exhibits significant advantages over the other \textit{Attention} variants.
These results, which positively answer our second research question, \ref{rq2}, highlight the importance of correctly managing the user-related information in personalization and the potential of deepening this research area.

\begin{table}[t]
    \centering
    \caption{
    Effectiveness of BM25 and those of the Personalized Results Re-Ranking Framework with different user models.
    $\ast$ and $\dagger$ denote significant improvements in a Bonferroni corrected Fisher’s randomization test with $p < 0.001$ over \textit{Mean} and over all the baselines, respectively.
    Best results are highlighted in boldface.
    }
    
    \begin{tabular}{l|l|ccc|cc}
    \toprule
    \toprule
    \multicolumn{7}{c}{\textbf{Web Search Dataset}} \\
    \midrule
    \textbf{Model}
    & \textbf{Alignment}
    & \textbf{MAP@100} & \textbf{MRR@10} & \textbf{NDCG@10}
    & $\bm{\lambda}$ & $\bm{\sigma(t)}$ \\
    \midrule
    BM25 & --- &
    0.245\hphantom{$^{\ast}$} & 
    0.238\hphantom{$^{\ast}$} & 
    0.280\hphantom{$^{\ast}$} & 
    --- & --- \\
    \midrule
    Mean & --- &
    0.282\hphantom{$^{\ast}$} &
    0.276\hphantom{$^{\ast}$} &
    0.329\hphantom{$^{\ast}$} &
    0.2 & --- \\
    \midrule
    \multirow{3}{*}{\makecell{Attention}} &
    Additive &
    0.281\hphantom{$^{\ast}$} &
    0.276\hphantom{$^{\ast}$} &
    0.328\hphantom{$^{\ast}$} &
    0.2 & --- \\
    & Cosine &
    0.287$^{\ast}$ &
    0.281$^{\ast}$ &
    0.335$^{\ast}$ &
    0.2 & --- \\
    & Scaled-Dot &
    0.290$^{\ast}$ &
    0.285$^{\ast}$ &
    0.339$^{\ast}$ &
    0.2 & --- \\
    \midrule
    \multirow{3}{*}{\makecell{Zero\\Attention}} &
    Additive &
    0.277\hphantom{$^{\ast}$} &
    0.272\hphantom{$^{\ast}$} &
    0.325\hphantom{$^{\ast}$} &
    0.2 & --- \\
    & Cosine &
    0.286$^{\ast}$ &
    0.281$^{\ast}$ &
    0.334$^{\ast}$ &
    0.2 & --- \\
    & Scaled-Dot &
    0.290$^{\ast}$ &
    0.285$^{\ast}$ &
    0.338$^{\ast}$ &
    0.2 & --- \\
    \midrule
    Multi-Head &
    Scaled-Dot &
    0.275\hphantom{$^{\ast}$} &
    0.269\hphantom{$^{\ast}$} &
    0.324\hphantom{$^{\ast}$} &
    0.2 & --- \\
    \midrule
    Denoising &
    Cosine-based &
    \textbf{0.338}$^{\dagger}$ &
    \textbf{0.336}$^{\dagger}$ &
    \textbf{0.393}$^{\dagger}$ &
    0.4 & 0.7 \\
    \midrule
    \multicolumn{7}{c}{\textbf{Academic Search Dataset}} \\
    \midrule
    \textbf{Model}
    & \textbf{Alignment}
    & \textbf{MAP@100} & \textbf{MRR@10} & \textbf{NDCG@10}
    & $\bm{\lambda}$ & $\bm{\sigma(t)}$ \\
    \midrule
    BM25 & --- &
    0.119\hphantom{$^{\ast}$} & 
    0.294\hphantom{$^{\ast}$} & 
    0.171\hphantom{$^{\ast}$} & 
    --- & --- \\
    \midrule
    Mean & --- &
    0.146\hphantom{$^{\ast}$} &
    0.328\hphantom{$^{\ast}$} &
    0.200\hphantom{$^{\ast}$} &
    0.6 & --- \\
    \midrule
    \multirow{3}{*}{\makecell{Attention}} &
    Additive &
    0.156$^{\ast}$ &
    0.340$^{\ast}$ &
    0.213$^{\ast}$ &
    0.6 & --- \\
    & Cosine &
    0.151\hphantom{$^{\ast}$} &
    0.332\hphantom{$^{\ast}$} &
    0.206\hphantom{$^{\ast}$} &
    0.6 & --- \\
    & Scaled-Dot &
    0.157$^{\ast}$ &
    0.343$^{\ast}$ &
    0.214$^{\ast}$ &
    0.6 & --- \\
    \midrule
    \multirow{3}{*}{\makecell{Zero\\Attention}} &
    Additive &
    0.155$^{\ast}$ &
    0.338\hphantom{$^{\ast}$} &
    0.211$^{\ast}$ &
    0.6 & --- \\
    & Cosine &
    0.150\hphantom{$^{\ast}$} &
    0.330\hphantom{$^{\ast}$} &
    0.204\hphantom{$^{\ast}$} &
    0.6 & --- \\
    & Scaled-Dot &
    0.156$^{\ast}$ &
    0.341$^{\ast}$ &
    0.212$^{\ast}$ &
    0.6 & --- \\
    \midrule
    Multi-Head &
    Scaled-Dot &
    0.152\hphantom{$^{\ast}$} &
    0.336\hphantom{$^{\ast}$} &
    0.207\hphantom{$^{\ast}$} &
    0.6 & --- \\
    \midrule
    Denoising &
    Cosine-based &
    \textbf{0.179}$^{\dagger}$ &
    \textbf{0.378}$^{\dagger}$ &
    \textbf{0.241}$^{\dagger}$ &
    0.6 & 0.6 \\
    \bottomrule
    \end{tabular}
    
    \label{tab:framework_results}
\end{table}

\subsection{Weighting Schemes Comparison}

In this section, aiming to answer our third research question (\ref{rq3}), we compare the mechanism employed by our proposed Attention variant to weigh the contribution of the user-related data in composing the user model at query time with the other considered Attention baselines.
We also consider the Mean-based user model, which uses an even weighting scheme, as a reference.
To conduct this evaluation, we consider only the queries for which Denoising Attention outputs a non-zero user model and employs only the scores deriving from the comparisons between the user models and the documents to re-rank the initially retrieved BM25 result lists.
We assume that if a user model achieves significantly better retrieval effectiveness than the others, then its weighting scheme is better, as we employ the same user-related information to build all the user models.
Since the goal of re-ranking is to improve over a first-stage retriever, we also consider BM25 results for reference.

As reported in Table \ref{tab:personalization_results}, there is generally a clear difference in the retrieval effectiveness of the \textit{Attention}-based user models and that of the \textit{Mean}-based user model, highlighting the potential of weighing the contribution of multiple sources of user-related information during personalization.
In the best case scenario (\textit{Scaled-dot}), the \textit{Attention}-based baselines increased over \textit{Mean} by 11\%, 13\%, and 11\% in MAP, MRR, and NDCG, respectively, on the \textit{Web Search Dataset}, and by 34\%, 30\%, and 34\% in MAP, MRR, and NDCG, respectively, on the \textit{Academic Search Dataset}.

Despite the already good improvements brought by the Attention as a mechanism for weighting user-related data in query-aware personalization, the Denoising Attention reached a significantly higher level of effectiveness.
With respect to the best performing Attention baselines, the Denoising Attention improved by 73\%, 86\%, and 76\% in MAP, MRR, and NDCG, respectively, on the \textit{Web Search Dataset}, and by 57\%, 53\%, and 55\% in MAP, MRR, and NDCG, respectively, on the \textit{Academic Search Dataset}.
This great difference is due to both the shortcoming of the Attention mechanism described in Section~\ref{shortcomings}, which makes it under-perform in many situations, and the solution we have proposed to solve them, which allows filtering noisy information and does not reduce the diversity of the estimated importance of the user-related documents.
We also highlight that the Denoising Attention-based user model is the only user model that improved over our first stage retriever, BM25, while of the other considered user models largely decreased its effectiveness.
Specifically, the \textit{Denoising Attention}-based user model improved over BM25 by 10\%, 10\%, and 14\% in MAP, MRR, and NDCG, respectively, on the \textit{Web Search Dataset}, and by 19\%, 08\%, and 13\% in MAP, MRR, and NDCG, respectively, on the \textit{Academic Search Dataset}.
These results positively answer our third research question, \ref{rq3}

\begin{table}[t]
    \centering
    \caption{
    Effectiveness of BM25 and those of the user models when used in isolation.
    $\ast$ and $\dagger$ denote significant improvements in a Bonferroni corrected Fisher’s randomization test with $p < 0.001$ over \textit{Mean} and over all the baselines, respectively.
    Best results are highlighted in boldface.
    }
    
    \begin{tabular}{l|l|ccc|cc}
    \toprule
    \toprule
    \multicolumn{7}{c}{\textbf{Web Search Dataset}} \\
    \midrule
    \textbf{Model}
    & \textbf{Alignment}
    & \textbf{MAP@100} & \textbf{MRR@10} & \textbf{NDCG@10}
    & $\bm{\lambda}$ & $\bm{\sigma(t)}$ \\
    \midrule
    BM25 & --- &
    0.240$^{\ast}$ & 
    0.233$^{\ast}$ & 
    0.274$^{\ast}$ & 
    --- & --- \\
    \midrule
    Mean & --- &
    0.136\hphantom{$^{\ast}$} &
    0.120\hphantom{$^{\ast}$} &
    0.157\hphantom{$^{\ast}$} &
    1.0 & --- \\
    \midrule
    \multirow{3}{*}{\makecell{Attention}} &
    Additive &
    0.136\hphantom{$^{\ast}$} &
    0.120\hphantom{$^{\ast}$} &
    0.155\hphantom{$^{\ast}$} &
    1.0 & --- \\
    & Cosine &
    0.141$^{\ast}$ &
    0.125$^{\ast}$ &
    0.166$^{\ast}$ &
    1.0 & --- \\
    & Scaled-Dot &
    0.152$^{\ast}$ &
    0.137$^{\ast}$ &
    0.177$^{\ast}$ &
    1.0 & --- \\
    \midrule
    \multirow{3}{*}{\makecell{Zero\\Attention}} &
    Additive &
    0.125\hphantom{$^{\ast}$} &
    0.108\hphantom{$^{\ast}$} &
    0.144\hphantom{$^{\ast}$} &
    1.0 & --- \\
    & Cosine &
    0.148$^{\ast}$ &
    0.132$^{\ast}$ &
    0.169$^{\ast}$ &
    1.0 & --- \\
    & Scaled-Dot &
    0.153$^{\ast}$ &
    0.138$^{\ast}$ &
    0.177$^{\ast}$ &
    1.0 & --- \\
    \midrule
    Multi-Head &
    Scaled-Dot &
    0.128\hphantom{$^{\ast}$} &
    0.111\hphantom{$^{\ast}$} &
    0.148\hphantom{$^{\ast}$} &
    1.0 & --- \\
    \midrule
    Denoising &
    Cosine-based &
    \textbf{0.264}$^{\dagger}$ &
    \textbf{0.256}$^{\dagger}$ &
    \textbf{0.312}$^{\dagger}$ &
    1.0 & 0.7 \\
    \midrule
    \multicolumn{7}{c}{\textbf{Academic Search Dataset}} \\
    \midrule
    \textbf{Model}
    & \textbf{Alignment}
    & \textbf{MAP@100} & \textbf{MRR@10} & \textbf{NDCG@10}
    & $\bm{\lambda}$ & $\bm{\sigma(t)}$ \\
    \midrule
    BM25 & --- &
    0.120$^{\ast}$ & 
    0.295$^{\ast}$ & 
    0.172$^{\ast}$ & 
    --- & --- \\
    \midrule
    Mean & --- &
    0.068\hphantom{$^{\ast}$} &
    0.160\hphantom{$^{\ast}$} &
    0.094\hphantom{$^{\ast}$} &
    1.0 & --- \\
    \midrule
    \multirow{3}{*}{\makecell{Attention}} &
    Additive &
    0.090$^{\ast}$ &
    0.205$^{\ast}$ &
    0.123$^{\ast}$ &
    1.0 & --- \\
    & Cosine &
    0.076$^{\ast}$ &
    0.172$^{\ast}$ &
    0.103$^{\ast}$ &
    1.0 & --- \\
    & Scaled-Dot &
    0.091$^{\ast}$ &
    0.208$^{\ast}$ &
    0.125$^{\ast}$ &
    1.0 & --- \\
    \midrule
    \multirow{3}{*}{\makecell{Zero\\Attention}} &
    Additive &
    0.086$^{\ast}$ &
    0.195$^{\ast}$ &
    0.117$^{\ast}$ &
    1.0 & --- \\
    & Cosine &
    0.075$^{\ast}$ &
    0.171$^{\ast}$ &
    0.103$^{\ast}$ &
    1.0 & --- \\
    & Scaled-Dot &
    0.088$^{\ast}$ &
    0.201$^{\ast}$ &
    0.120$^{\ast}$ &
    1.0 & --- \\
    \midrule
    Multi-Head &
    Scaled-Dot &
    0.074$^{\ast}$ &
    0.172$^{\ast}$ &
    0.101$^{\ast}$ &
    1.0 & --- \\
    \midrule
    Denoising &
    Cosine-based &
    \textbf{0.143}$^{\dagger}$ &
    \textbf{0.319}$^{\dagger}$ &
    \textbf{0.194}$^{\dagger}$ &
    1.0 & 0.6 \\
    \bottomrule
    \end{tabular}
    
    \label{tab:personalization_results}
\end{table}

\subsection{Robustness}

In this section, we evaluate and discuss the \textit{robustness} of the considered user models when used in combination with BM25, aiming to answer our fourth research question, \ref{rq4}.
Specifically, we consider the number of times personalization decreased BM25 effectiveness in terms of MAP@100.
We remind the reader that the Web Search Dataset based on the AOL query log we employed has 36\,052 test queries, while the Academic Search Dataset has 24\,056.
Table \ref{tab:robusteness} shows the number of times personalization was actually harmful to the retrieval effectiveness in terms of MAP@100.
Quite surprisingly, the \textit{Attention}-based user models are often more harmful than the user model based on the average of the user document representations, Mean, although more effective in general, as previously reported.
Conversely, the \textit{Denoising Attention}-based user model considerably decreased for both datasets the number of times personalization harmed the retrieval process w.r.t. the other considered user models.
Compared to the \textit{Denoising Attention}-based user model, the best baselines on the Web Search Dataset and the Academic Search Dataset decreased the retrieval effectiveness of BM25 for 38\% and 8\% more queries, respectively.
The much more significant difference between the \textit{Denoising Attention}-based user model and the other considered user models on the Web Search Dataset than on the Academic Search Dataset is due to the different nature of the employed datasets.
In the Web Search Dataset, the user-related data accounts for many different user interests, while on the Academic Search Dataset, they are much more focused on particular topics.
In the first case, personalization is much more likely to harm the retrieval process if a filtering mechanism for the user information is not employed, as in the case of all the considered user models but the one based on our \textit{Attention} variant.
We conclude the \textit{Denoising Attention}-based user model is much more robust than the other considered user models regardless of the search scenario, which positively answer our fourth research question, \ref{rq4}.

\begin{table}[t]
    \centering
    \caption{Number of times (and ratios) personalization decreased BM25 effectiveness in terms of MAP@100 (lower is better).
    Best results are highlighted in boldface.
    Best baselines are highlighted in italic.}
    \begin{tabular}{l|l|c|c}
    \toprule
    \toprule
    \textbf{Model}
    & \textbf{Alignment}
    & \textbf{Web Search Dataset}
    & \textbf{Academic Search Dataset} \\
    \midrule
    Mean & --- &
    10\,798 (30\%) & 6\,165 (26\%) \\
    \midrule
    \multirow{3}{*}{\makecell{Attention}} &
    Additive &
    11\,157 (31\%) & 6\,076 (25\%) \\
    & Cosine &
    \phantom{1}9\,877 (27\%) & 6\,580 (27\%) \\
    & Scaled-Dot &
    \phantom{1}9\,426 (26\%) & \textit{5\,954 (25\%)} \\
    \midrule
    \multirow{3}{*}{\makecell{Zero\\Attention}} &
    Additive &
    11\,508 (32\%) & 6\,201 (26\%) \\
    & Cosine &
    10\,234 (28\%) & 6\,708 (28\%) \\
    & Scaled-Dot &
    \textit{\phantom{1}9\,356 (26\%)} & 6\,131 (25\%) \\
    \midrule
    Multi-Head &
    Scaled-Dot &
    12\,049 (33\%) & 6\,366 (26\%) \\
    \midrule
    Denoising &
    Cosine-based &
    \textbf{\phantom{1}6\,780 (19\%)} & \textbf{5\,509 (23\%)} \\
    \bottomrule
    \end{tabular}
    \label{tab:robusteness}
\end{table}

\subsection{Model Analysis}
\label{sub:model_analysis}
In this section, we first evaluate the \textit{Denoising Attention}-based user model performances for various threshold values.
Then, we compare the performances of the considered user models for queries with various amounts of associated user-related documents.
Finally, we ablate the design choices underlying our proposed \textit{Attention} variant.

\paragraph{\textbf{Threshold}}

\begin{figure*}[t]
    \centering
    \subfloat[\centering Retrieval effectiveness of the personalized re-ranking pipeline with \textit{Denoising Attention}-based user model on the \textit{Web Search Dataset}]{
        \resizebox{.39\linewidth}{!}{\includegraphics{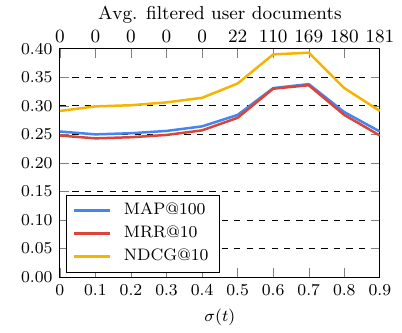} \label{fig:threshold_analysis_c}}
    }
    \hspace{0.05\textwidth}
    \subfloat[\centering Retrieval effectiveness of the personalized re-ranking pipeline with \textit{Denoising Attention}-based user model on the \textit{Academic Search Dataset}]{
        \resizebox{.39\linewidth}{!}{\includegraphics{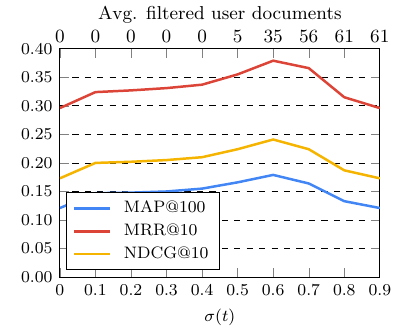} \label{fig:threshold_analysis_d}}
    } \\
    \subfloat[\centering Remaining user documents for $\bm{\sigma(t) = 0.7}$  on the \textit{Web Search Dataset}]{
        \resizebox{.4\linewidth}{!}{\includegraphics{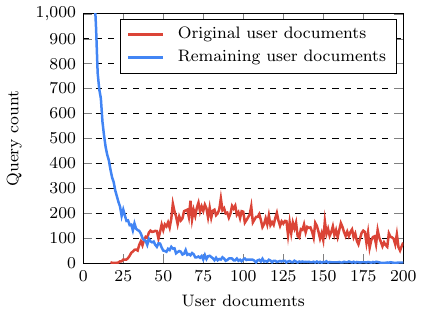}
         \label{fig:threshold_analysis_e}}
    }
    \hspace{0.05\textwidth}
    \subfloat[\centering Remaining user documents for $\bm{\sigma(t) = 0.6}$ on the \textit{Academic Search Dataset}]{
        \resizebox{.39\linewidth}{!}{\includegraphics{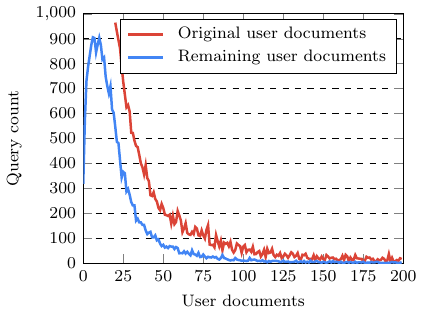}
         \label{fig:threshold_analysis_f}}
    } \\
    \caption{Threshold analysis.}
    \label{fig:threshold_analysis}
\end{figure*}

Figures \ref{fig:threshold_analysis_c} and \ref{fig:threshold_analysis_d} show the performances of the results re-ranking pipeline with the \textit{Denoising Attention}-based user model for different threshold values on the considered datasets.
The figures also report the average number of filtered user documents for each considered threshold value.
On average, the test queries have 181 and 61 associate user-related documents in the \textit{Web Search Dataset} and the \textit{Academic Search Dataset}, respectively, while the average number of filtered ones for the best threshold values are 169 and 35, respectively.
The different ratios of average filtered user-related documents are again due to the distinct nature of the two search scenarios and datasets.
Our proposed approach is able to adapt to different search contexts thanks to the threshold parameter and our filtering mechanism.
When the threshold is zero, which corresponds to not filtering any user-related document in our case, the model effectiveness is very low for both datasets.
When the threshold is equal to $0.5$, which corresponds to using the \textit{cosine similarity} with no modification as our \textit{alignment model}, the model still does not reach its full potential.
These results highlight again the need for a filtering mechanism that can be tuned and modulated.
In Figures \ref{fig:threshold_analysis_e} and \ref{fig:threshold_analysis_f}, we can observe how the distribution of the documents used for personalization changes using the \textit{Denoising Attention}-based user model.
As expected, the number of documents used for personalization decreases drastically in both the considered search scenarios.
In particular, for the \textit{Web Search} scenario, which comprises, for each user, a very diverse set of user-related documents, we registered a very pronounced selection of the user data used for personalization.
As the filtering mechanism employed by the \textit{Denoising Attention} is applied to each user document independently, the model can filter numerous documents that are not strictly related to the query.

\paragraph{\textbf{User Document Count}}

\begin{figure*}[t]
    \centering
    \subfloat[\centering \textit{Web Search Dataset}]{
        \resizebox{0.9\linewidth}{!}{\includegraphics{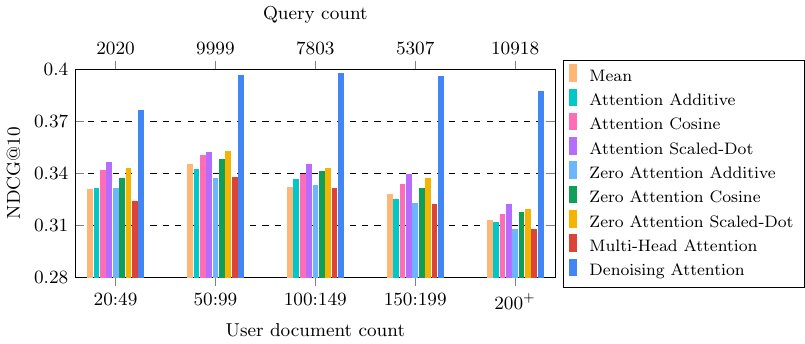} \label{fig:user_document_count_a}}
    } \\
    \subfloat[\centering \textit{Academic Search Dataset}]{
        \resizebox{0.9\linewidth}{!}{\includegraphics{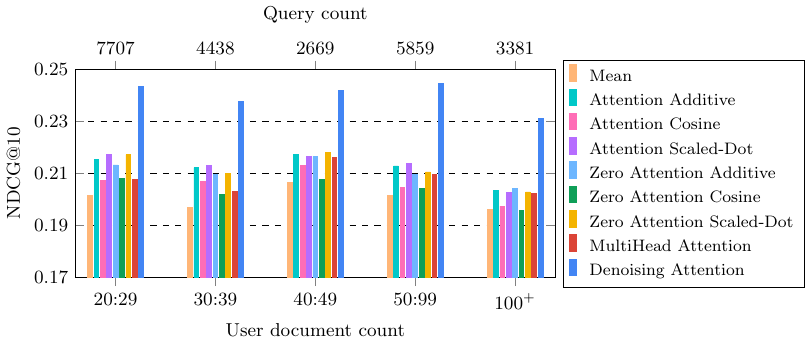} \label{fig:user_document_count_b}}
    }
    \caption{Effectiveness of the user models when combined with BM25 for queries with different amounts of associated user-related documents.}
    \label{fig:user_document_count}
\end{figure*}

Figures \ref{fig:user_document_count_a} and \ref{fig:user_document_count_b} report the performances of the results re-ranking pipeline with each of the considered user models on the test queries grouped by their amounts of associated user-related documents on the \textit{Web Search Dataset} and the \textit{Academic Search Dataset}, respectively.
We grouped queries by the range in which their associated amount of user documents lays.
In particular, we considered the queries in the following ranges: 
20:49, 50:99, 100:149, 150:199, and $200^{+}$ for the \textit{Web Search Dataset}, and 20:29, 30:39, 40:49, 50:59, and $100^{+}$ for the \textit{Academic Search Dataset}.
We chose different ranges for the two datasets as their user document distribution is very different, as shown in Figures \ref{fig:threshold_analysis_c} and \ref{fig:threshold_analysis_d}.
Moreover, by datasets construction (Section~\ref{dataset}), all queries have at least 20 associated user documents.
As shown in the figures, the baselines achieved mixed results.
It is not clear at a glance which one performed the best overall.
Looking closely, we can see that the \textit{Attention}-based user model with the \textit{scaled-dot alignment model} (\textit{purple bar}) generally performed better than or equal to the other baselines in the considered ranges.
Conversely, the \textit{Denoising Attention}-based user model consistently outperformed all the considered baselines for each group of queries in each dataset, showing its benefits generalize regardless of the number of available user documents and search scenario.

\paragraph{\textbf{Ablation Study}}

Table \ref{tab:ablation_results} shows the performances of the results re-ranking pipeline with the \textit{Denoising Attention}-based user model and with some variations derived by ablating our proposal. For comparison purposes, we added the results of the best performing baseline from previous experiments, the \textit{Attention}-based model with the \textit{scaled-dot alignment model}.
The first variation of our proposed \textit{Attention} variant, called \textit{Filter Attention}, employs the \textit{ReLU}-based filtering mechanism we proposed and Eq.~\ref{eq:denoising_norm} for the \textit{normalization} step to the \textit{Attention} with the \textit{scaled-dot alignment model}.
As discussed before, using Eq.~\ref{eq:denoising_norm} for the \textit{normalization} step is mandatory for the filtering mechanism to have effect as \textit{Softmax} normalization is \textit{translationally invariant}.
We did not use the threshold parameter in this variation of the \textit{Denoising Attention} as the employed \textit{alignment models}' output is unbounded, making it difficult to calibrate such a parameter.
For the other variation we considered the \textit{Denoising Attention} with \textit{Softmax} normalization to show the need of using Eq.~\ref{eq:denoising_norm} and departing from \textit{Softmax} to make our proposed mechanism to work properly. We called this variation \textit{Denoising Softmax}. 

As shown in the table, \textit{Denoising Softmax} performs poorly and decreases the retrieval effectiveness of the system w.r.t. our best performing baseline on both datasets, confirming that using our filtering mechanism with \textit{Softmax} normalization does not perform properly.
On the other hand, \textit{Filter Attention} significantly improved over the \textit{Attention}-based model with the \textit{scaled-dot alignment model}, corroborating our intuition that the \textit{Softmax} normalization is not optimal in the context of personalization and suggesting the proposed alternative is effective regardless of the employed \textit{alignment model}.
The \textit{Denoising Attention} significantly outperformed all of its considered variations, verifying the utility of our design choices and their complementarity.

\begin{table}[t]
    \centering
    \caption{
    Effectiveness of the Personalized Results Re-Ranking Framework with different \textit{Denoising Attention} variations.
    $\dagger$ denotes significant improvements in a Bonferroni corrected Fisher’s randomization test with $p < 0.001$ over over all the baselines.
    Best results are highlighted in boldface.
    }
    
    \begin{tabular}{l|l|ccc|cc}
    \toprule
    \toprule
    \multicolumn{7}{c}{\textbf{Web Search Dataset}} \\
    \midrule
    \textbf{Model}
    & \textbf{Alignment}
    & \textbf{MAP@100} & \textbf{MRR@10} & \textbf{NDCG@10}
    & $\bm{\lambda}$ & $\bm{\sigma(t)}$ \\
    \midrule
    Attention & Scaled-Dot &
    0.290\hphantom{$^{\dagger}$} &
    0.285\hphantom{$^{\dagger}$} &
    0.339\hphantom{$^{\dagger}$} &
    0.2 & --- \\
    \midrule
    Filter Attention & Scaled-Dot &
    0.299\hphantom{$^{\dagger}$} &
    0.294\hphantom{$^{\dagger}$} &
    0.351\hphantom{$^{\dagger}$} &
    0.3 & --- \\
    \midrule
    Denoising Softmax &
    Cosine-based &
    0.285\hphantom{$^{\dagger}$} &
    0.280\hphantom{$^{\dagger}$} &
    0.334\hphantom{$^{\dagger}$} &
    0.2 & 0.1 \\
    \midrule
    Denoising &
    Cosine-based &
    \textbf{0.338}$^{\dagger}$ &
    \textbf{0.336}$^{\dagger}$ &
    \textbf{0.393}$^{\dagger}$ &
    0.4 & 0.7 \\
    \midrule
    \multicolumn{7}{c}{\textbf{Academic Search Dataset}} \\
    \midrule
    \textbf{Model}
    & \textbf{Alignment}
    & \textbf{MAP@100} & \textbf{MRR@10} & \textbf{NDCG@10}
    & $\bm{\lambda}$ & $\bm{\sigma(t)}$ \\
    \midrule
    Attention & Scaled-Dot &
    0.157\hphantom{$^{\dagger}$} &
    0.343\hphantom{$^{\dagger}$} &
    0.214\hphantom{$^{\dagger}$} &
    0.6 & --- \\
    \midrule
    Filter Attention & Scaled-Dot &
    0.165\hphantom{$^{\dagger}$} &
    0.354\hphantom{$^{\dagger}$} &
    0.223\hphantom{$^{\dagger}$} &
    0.6 & --- \\
    \midrule
    Denoising Softmax &
    Cosine-based &
    0.151\hphantom{$^{\dagger}$} &
    0.332\hphantom{$^{\dagger}$} &
    0.206\hphantom{$^{\dagger}$} &
    0.5 & 0.1 \\
    \midrule
    Denoising &
    Cosine-based &
    \textbf{0.180}$^{\dagger}$ &
    \textbf{0.382}$^{\dagger}$ &
    \textbf{0.243}$^{\dagger}$ &
    0.6 & 0.6 \\
    \bottomrule
    \end{tabular}
    
    \label{tab:ablation_results}
\end{table}

\section{Conclusion and Future Works}

In this work, we have addressed some issues related to the use of the \textit{Attention} mechanism for query-aware user modeling and proposed a novel user-data aggregation model called \textit{Denoising Attention}, designed to solve the shortcomings of the standard \textit{Attention} formulation and, in particular, filter out noisy user-related information.
Experimental evaluation in two different search scenarios, namely \textit{Web Search} and \textit{Academic Search}, shows the benefits of our proposed approach over other \textit{Attention} variants and highlights the potential of correctly managing the user-related information.
Finally, the ablation study we conducted clearly illustrates the benefits of our design choices and their synergy.
Despite the significant improvements brought by our proposed Denoising Attention mechanism when applied for selecting user-related information for query-aware personalization, some related problems are worth further study.
The alignment model we employed, the scaled cosine similarity, could be replaced by a parameterized function that could leverage additional information other than the textual-based representations of a user-related document and the query.
For example, the dates associated with the user-related documents might play a role in personalization, as documents written or consulted long before the query might be less relevant to personalization than more recent ones, despite being semantically related to the current search.
Furthermore, the fixed value threshold parameter we employed could be sub-optimal in many cases. 
As shown by the difference in the threshold parameter values for the two considered datasets, different queries could benefit from more user-related information or require a finer selection of the user-related data employed in the personalization process.
To conclude, the management of the user-related information during personalization is fundamental and far from being a solved issue, leaving room for further improvements.

\bibliographystyle{ACM-Reference-Format}
\bibliography{bibliography}

\end{document}